\documentclass[11pt]{article} 
\usepackage{amsfonts,amscd,amsmath,amssymb,graphicx}
\newcommand{\oh}{\frac{1}{2}}
\def\4{\tfrac{1}{4}}
\def\ep{\text{e}}

\def\g{\mathfrak{g}}

\def\z{z_{\text{\tiny 0}}}
\def\zt{z_{\text{\tiny T}}}
\def\zc{z_{\text{\tiny c}}}
\def\T{T_c}

\title{The Spatial String Tension, Thermal Phase Transition, and AdS/QCD}
\author{Oleg Andreev\thanks{andre@itp.ac.ru}\\\\
{\it L.D. Landau Institute for Theoretical Physics, Kosygina 2, 119334 Moscow, Russia}\\\\
Valentine I. Zakharov\thanks{xxz@mppmu.mpg.de}\\\\
{\it Istituto Nazionale di Fisica Nucleare -- Sezione di Pisa} \\
{\it Dipartimento di Fisica Universita di Pisa, Largo Pontecorvo 3, 56127 Pisa, Italy;}\\
{\it Max-Planck Institut f\" ur Physik, F\" ohringer Ring 6, 80805 M\" unchen, Germany} }
\date{}

\begin{document} 

\maketitle 
\begin{abstract} 
We present results of modeling the temperature dependence of the spatial string tension and thermal phase transition in 
a five-dimensional framework nowadays known as AdS/QCD. For temperatures close to the critical one we find a behaviour 
remarkably consistent with the lattice results.
\\
{\it PACS:} 12.38.Lg; 12.90.+b\\
{\it Keywords:} Spatial string tension; Thermal phase transition; AdS/QCD
 \end{abstract}

\vspace{-13.5cm}
\begin{flushright}
LANDAU-06-A2
\end{flushright}
\vspace{12cm}

\section{Introduction}
It is well known that $SU(N)$ gauge theories at high temperature undergo a phase transition to a deconfined phase. Although 
at the phase transition point basic thermodynamic observables show a drastic qualitative change, there are some correlation 
functions of physical interest whose structure does not change qualitatively at $\T$. For example, this is the case for 
the pseudo-potential extracted from spatial Wilson loops. It is confining for all temperatures \cite{pseudo}. This is 
taken as indication that certain confining properties survive in the high temperature phase. 
While the leading high temperature behaviour of the pseudo-potential for temperatures well 
above $\T$ can be understood in terms of high temperature perturbation theory, 
non-perturbative effects make it difficult to compute that near the phase transition point. So far only a lattice study of the 
whole temperature dependence of the spatial string tension exists \cite{bali, karsch}.

The situation changed drastically with the invention 
of the AdS/CFT correspondence that resumed interest in finding a string description 
of strong interactions. Its more phenomenological cousin called AdS/QCD  
deals with a five-dimensional effective description and tries to fit it to 
QCD as much as possible. 

In this Letter we will explore the temperature dependence of the 
spatial string tension and thermal phase transition within AdS/QCD. 

There are good motivations for doing so. 
First, in a recent paper \cite{az} we have constructed a phenomenological heavy-quark potential which 
has a remarkable similarity to the Cornell potential. 
We did so by using the slightly deformed AdS metric. 
Second, it is already known that 
the thermodynamics of ${\cal N}=4$ super Yang-Mills 
theory is related to the AdS black hole geometry \cite{witt}. 
What we need now is to 
include the thermodynamics into the five-dimensional framework of \cite{az}. 
To this end, we consider the following Euclidean background 
metric

\begin{equation}\label{metric}
ds^2=
R^2\frac{h}{z^2}
\left(f dt^2+dx^2_i+\frac{1}{f}dz^2\right)
\,,\quad\quad 
h(z)=\ep^{\oh cz^2}
\,,\quad\quad
f(z)=1-\bigl(\tfrac{z}{\zt}\bigr)^4
\,,
\end{equation}
where $i=1,2 ,3$. $t$ is a periodic variable of period $\pi\zt$. 

At zero value of the deformation parameter $c$ we have the $\text{AdS}_5$ 
black hole metric, as expected. Here $\zt$ is related to the Hawking temperature 
\begin{equation}\label{T}
T=\frac{1}{\pi\zt}
\,
\end{equation}
whose dual description is nothing but the temperature of gauge theory. 
On the other hand, at $T=0$ we have in fact the slightly 
deformed $\text{AdS}_5$ metric. Such a deformation is notable. 
The point is that in this background linearized five-dimensional Yang-Mills 
equations are effectively reduced to Laguerre differential equation. 
As a result, the spectrum turns out to be like that of the linear 
Regge models \cite{mets,oa}. 
This fact allows us to fix the value of $c$ from the $\rho$ meson trajectory. It is of order \cite{oa}

\begin{equation}\label{c}
c\approx 0.9\,\text{GeV\,}^2
\,.
\end{equation}
The central point now is that the metric \eqref{metric} {\it does not} contain 
any free fit parameter. Thus, evaluation of the spatial string tension which we will
undertake can be considered as a crucial test of AdS/QCD in the infrared region.
\section{Calculating the Spatial String Tension}

Given the background metric,  we can calculate expectation values of Wilson loops by using the standard formalism of 
AdS/CFT.\footnote{The literature on the Wilson loops at finite temperature within the AdS/CFT 
correspondence is vast. For a discussion of this issue, see, e.g., \cite{Tloops} and references therein.} 
The Wilson loops obeying an 
area law provide string tensions. Our goal is therefore to study spatial Wilson loops.

To this end, we consider a rectangular loop $\cal{C}$ along two spatial directions $(x,y)$ on the boundary ($z=0$) of 
five-dimensional space. As usual, we take one direction to be large, say $Y\rightarrow\infty$. 
The quark and antiquark are 
set at $x=\tfrac{r}{2}$ and $x=-\tfrac{r}{2}$, respectively. 

Next, we make use of the Nambu-Goto action equip with the background metric \eqref{metric} and choose the world-sheet 
coordinates as $\xi^1=x$ and $\xi^2=y$. This yields

\begin{equation}\label{ng1}
S=\frac{\g }{2\pi}Y\int^{\,\tfrac{r}{2}}_{-\tfrac{r}{2}} 
dx\,\frac{h}{z^2}
\sqrt{1+\frac{1}{f}(z')^2}
\,,
\end{equation}
where $\g=\tfrac{R^2}{\alpha'}$. A prime denotes a derivative with respect to $x$.

Now it is easy to find the equation of motion for $z$
\begin{equation}\label{eqm}
zz''+\left(f+(z')^2\right)\left(2-z\partial_z\ln h\right)-\oh z(z')^2\partial_z\ln f=0
\end{equation}
as well as the first integral
\begin{equation}\label{int}
\frac{h}{z^2\sqrt{1+\frac{1}{f}(z')^2}}=C
\,.
\end{equation}
The integration constant $C$ can be expressed via the maximum value of $z$. On symmetry grounds, $z$ reaches it at 
$x=0$.\footnote{Note that Eq.\eqref{int} has two solutions. One has $\zt=z\vert_{x=0}$. The solution of interest has 
$C=\tfrac{h}{z^2}\vert_{x=0}$.} By virtue of \eqref{int}, the integral over $\left[-\tfrac{r}{2},\tfrac{r}{2}\right]$ of $dx$ is equal to 

\begin{equation}\label{r}
r=2 \z
\int^1_0 dv\,v^2
\exp\bigl\lbrace\bigl(\tfrac{\z}{\zc}\bigr)^2(1-v^2)\bigr\rbrace
\Bigl(1-\bigl(\tfrac{\z}{\zt}\bigr)^4 v^4\Bigr)^{-\tfrac{1}{2}}
\Bigl(1-v^4
\exp\bigl\lbrace2\bigl(\tfrac{\z}{\zc}\bigr)^2(1-v^2)\bigr\rbrace
\Bigr)^{-\tfrac{1}{2}}
\,,
\end{equation}
where $v=\tfrac{z}{z_0}$, $\zc=\sqrt{\tfrac{2}{c}}$, and $\z=z\vert_{x=0}$.

At this point a comment is in order. A simple analysis shows that the integral \eqref{r} is real for $\z$ subject to 

\begin{subequations}\label{walls}
\begin{gather}
\z<\zt \,, \label{wt}\\
\z<\zc 
\,. \label{wc}
\end{gather}
\end{subequations}
Note that in the limit as $c$ goes to zero $\z$ is bounded by a horizon ($z=\zt$), as should be for the black hole geometry. This gives 
rise to  the first wall \eqref{wt}. On the other hand, for zero temperature it is also bounded if $c\not= 0$. The physical reason for 
this is a gravitational force appeared because of  the deformation of the 
AdS metric.\footnote{O.A. thanks L. Susskind for a discussion of this issue.} The 
easiest way to see what is going on is to introduce the effective string tension depending on $z$. It is simply 
$\sigma(z)=z^{-2}\exp\lbrace\tfrac{1}{2}cz^2\rbrace$ as follows from the form of the metric. Now consider the behavior of 
a string bit in the potential $V=\sigma(z)$ shown in Fig.1. 
%
\begin{figure}[ht]
\begin{center}
\includegraphics[width=4.75cm]{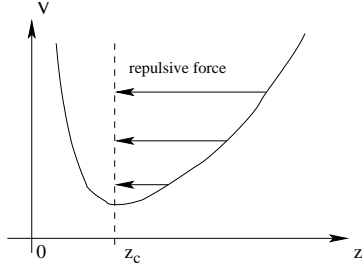}
\caption{\small{Schematic representation of the potential.}}
\end{center}
\end{figure}
The potential reaches its minimum value exactly at $z=\zc$, so a repulsive 
force prevents the string from getting deeper in $z$ direction. Note that because the string ends on the (infinitely) heavy 
quark-antiquark pair set at $z=0$, it doesn't completely roll down to the minima of the potential. This gives rise to the second 
wall \eqref{wc} of \cite{az}. In summary, there are the two walls in the problem in question. This fact will play a key role in 
understanding the temperature dependence of the spatial string tension.
 
Now, as in \cite{az}, we will compute the energy of the configuration. In the process we regularize the integral over $z$ by imposing a 
cutoff $\epsilon$. Finally, the regularized expression takes the form

\begin{equation}\label{energy}
E_{\text{\tiny R}}=\frac{\g}{\pi\z}
\int^1_{\tfrac{\epsilon}{z_0}}dv\,v^{-2}
\exp\bigl\lbrace\bigl(\tfrac{\z}{\zc}\bigr)^2 v^2\bigr\rbrace
\Bigl(1-\bigl(\tfrac{\z}{\zt}\bigr)^4 v^4\Bigr)^{-\tfrac{1}{2}}
\Bigl(1-v^4\exp\bigl\lbrace 2\bigl(\tfrac{\z}{\zc}\bigr)^2(1-v^2)\bigr\rbrace\Bigr)^{-\tfrac{1}{2}}
\,.
\end{equation}
Its $\epsilon$-expansion is simply 

\begin{equation}\label{energy1}
E_{\text{\tiny R}}=\frac{\g}{\pi\epsilon}+E+O(\epsilon)
\,,
\end{equation}
where 

\begin{equation}\label{E}
E=\frac{\g}{\pi\z}
\biggl(-1+
\int^1_0 \frac{dv}{v^2}\Bigl[
\exp\bigl\lbrace\bigl(\tfrac{\z}{\zc}\bigr)^2 v^2\bigr\rbrace
\Bigl(1-\bigl(\tfrac{\z}{\zt}\bigr)^4 v^4\Bigr)^{-\tfrac{1}{2}}
\Bigl(1-v^4\exp\bigl\lbrace2\bigl(\tfrac{\z}{\zc}\bigr)^2(1-v^2)\bigr\rbrace\Bigr)^{-\tfrac{1}{2}}
-1
\Bigr]
\biggr)
\,.
\end{equation}

\vspace{0.2cm}
\noindent Similarly as $r$, $E$ is real only for $\z$ subject to the constraints \eqref{walls}.

As in the case of zero temperature \cite{az},  the pseudo-potential in question is written in parametric form given by Eqs.\eqref{r} and 
\eqref{E}. It is unclear to us how to eliminate the parameter $\z$ and find $E$ as a function of $r$, $T$, and $c$. We 
can, however, gain some important insights from numerical calculations. 

In Fig.2 we have plotted $E/\g$ against $r$. Apparently the pseudo-potential shows temperature dependence of its slope. Moreover, 
there exists a critical value of T such that the spatial string tension is temperature independent below $\T$ and rises rapidly 
above.\footnote{There is a subtle point here. As noted in \cite{az}, in the phenomenologically important 
interval $0.1\,\text{fm}\leq r\leq 1\,\text{fm}$ the slope of the potential is given by $\sigma$ or $\sigma_{\text{\tiny 0}}$ (coefficients 
in front of the linear terms at large and small distances) depending on the value of $c$. However, their ratio is of order $1.24$. Since 
it is not significant for our phenomenological estimates, we will be interested in $\sigma$ in what follows. } Interestingly enough, on 
the lattice such a picture was discovered in \cite{bali}. 
%
\begin{figure}[ht]
\begin{center}
\includegraphics[width=7.75cm]{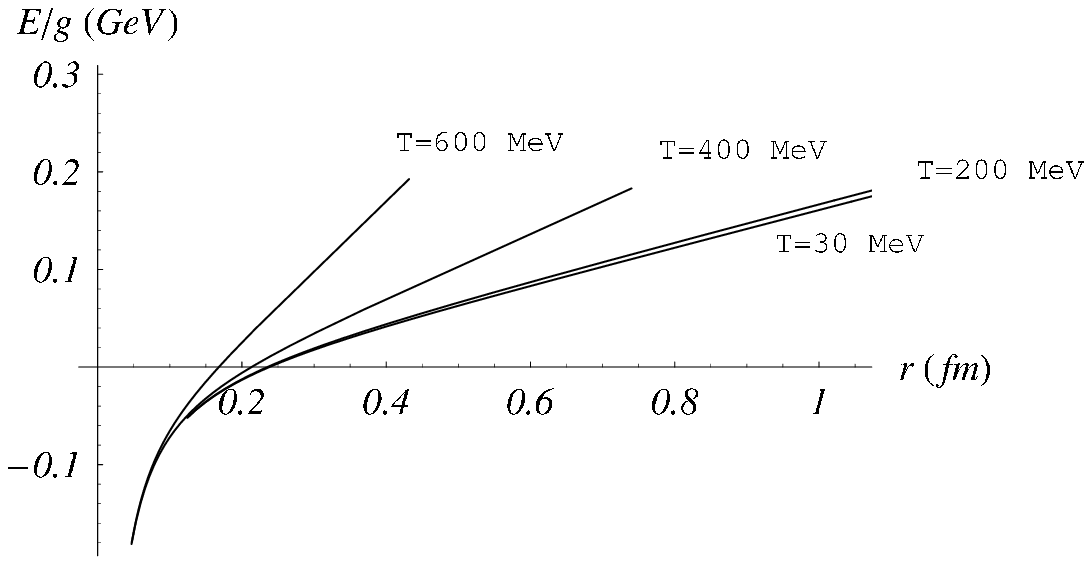}
\caption{\small{$E/\g$ as a function of $r$ at $T=30,\,200,\,400,\,600$ MeV, and $c=0.9\,\text{GeV}^2$.}}
\end{center}
\end{figure}

Having seen the pattern from the numerics, we can now try to  find a temperature dependence of the spatial string tension. 

First, let us have a close look at Eq.\eqref{r}. After a short inspection we find that $r$ is a continuously growing function of $\z$. This 
implies that large distances correspond to a region near the upper endpoint which is the smallest of $\zc$ and $\zt$. The integral is 
dominated by $v\sim 1$, where it takes the form $2\z\int_0^1 dv /\sqrt{a(1-v)+b(1-v)^2}$. Such an integral may be found in 
tables \cite{gr}. Finally, we get

\begin{equation}\label{r-large}
r=-\frac{2\z}{\sqrt{\beta}}\ln\Bigl(1-\frac{\z}{\zc}\Bigr)\Bigl(1-\frac{\z}{\zt}\Bigr) +O(1)
\,,
\end{equation}
where $\beta$ is a polynomial in $x=\bigl(\tfrac{\z}{\zt}\bigr)^4$ and $y=\bigl(\tfrac{\z}{\zc}\bigr)^2$. Explicitly, it is given by
\begin{equation}\label{beta}
\beta=-6+22x+18y-8y^2-34xy+8xy^2
\,.
\end{equation}
At the upper endpoint $r\rightarrow\infty$, as expected.

In a similar spirit, we can explore the long distance behavior of $E$. It follows from \eqref{E} that in the neighbor 
of the upper endpoint the energy behaves as  

\begin{equation}\label{E-large}
E=-\frac{\g\,\ep^{\bigl(\tfrac{\z}{\zc}\bigr)^2}}{\pi\z\sqrt{\beta}}
\ln\Bigl(1-\frac{\z}{\zc}\Bigr)\Bigl(1-\frac{\z}{\zt}\Bigr) 
+O(1)
\,.
\end{equation}
Along with the relation \eqref{r-large}, this means that at long distances the pseudo-potential is linear. The spatial 
string tension is given by 

\begin{equation}\label{tension-s}
\sigma_s=
\begin{cases}
\sigma & \text{if} \quad T\leq\T\,,\\
\sigma\bigl(\tfrac{T}{\T}\bigr)^2 \exp\bigl\lbrace \bigl(\tfrac{\T}{T}\bigr)^2-1\bigr\rbrace 
& \text{if} \quad T\geq\T\,.
\end{cases}
 \end{equation}
At this stage, we set $\sigma=\tfrac{\g\,\ep}{4\pi}c$ and $\T=\tfrac{1}{\pi}\sqrt{\tfrac{c}{2}}$. Note that $\sigma$ is the physical 
string tension at zero temperature \cite{az}.

\section{Thermal Phase Transition}
Now we will discuss the issue concerning the phase transition to the deconfined phase. 

We begin by making some qualitative comments about the temperature dependence of the spatial string tension. From the 
AdS/QCD perspective there are three possibilities shown in Fig.3. 
%
\begin{figure}[ht]
\begin{center}
\includegraphics[width=12 cm]{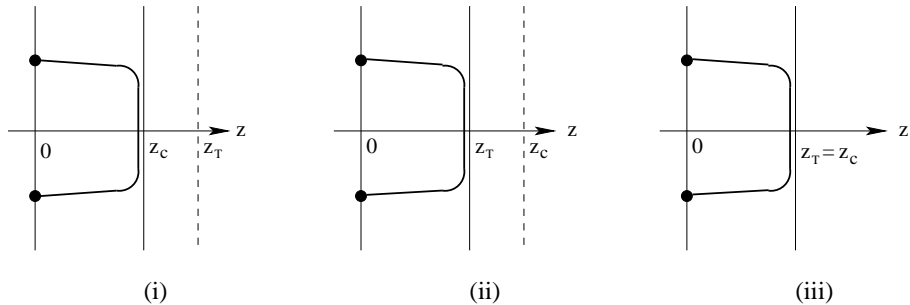}
\caption{\small{Schematic representation of the AdS/QCD pattern of the phase transition at $T=\T$.
(i) A low temperature phase. (ii) A high temperature phase. (iii) A point of the phase transition.
}}
\end{center}
\end{figure}
\newline (i) $\zc <\zt$. This means that the second wall \eqref{wc} terminates the string. If so, then the large distance physics of 
the string is determined by this wall. As a result, we find the same behavior as in \cite{az}. We therefore interpret this as the low 
temperature phase or, equivalently, the confined phase. 
\newline (ii) $\zc >\zt$. This time the first wall \eqref{wt} terminates the string. The large distance physics is determined by 
the near horizon geometry of the $\text{AdS}_5$ black hole and the spatial string tension scales like $T^2$ at high 
temperature \cite{witt}. We interpret this phase as the high temperature phase or, equivalently, the deconfined phase.
\newline (iii) $\zc=\zt$. This implies that the two walls coincide. We interpret this as a dual description (in the holographic 
sense of AdS/QCD) of the phase transition point. Put in a slightly different way, $\zc=\zt$, in terms of a critical temperature and 
the parameter $c$, is given by

\begin{equation}\label{Tc}
\T=\frac{1}{\pi}\sqrt{\frac{c}{2}}
\,.
\end{equation}

Having the AdS/QCD pattern of the phase transition, we can now make a couple of estimates relevant to phenomenology. 

It is of physical interest to estimate the critical temperature of our model. From \eqref{c} and \eqref{Tc} we have

\begin{equation}\label{Tc-MeV}
\T\approx 210\,\text{MeV}
\,.
\end{equation}

Next, we can estimate the ratio $\tfrac{\T}{\sqrt{\sigma}}$, where $\sigma$ is the zero temperature string tension. It is found by 
using Eqs.\eqref{tension-s} and \eqref{Tc}, 

\begin{equation}\label{normalization}
\frac{\T}{\sqrt{\sigma}}=\sqrt{\frac{2}{\ep\pi\g}}\approx 0.50
\,.
\end{equation}
The constant $\g$ was fixed in \cite{az} from the linear term of the Cornell potential. Its numerical 
value is of order $\g\approx 0.94$. The estimate \eqref{normalization} coincides within 7 per cent  with the lattice data 
for $SU(3)$ gauge theory \cite{karsch}. Thus, the agreement of theory with the data is very pleasing at this point.

Finally, we can compare the temperature dependence of the 
spatial string tension with the results of \cite{bali} for the high 
temperature phase of $SU(2)$ gauge theory. From the fit $\tfrac{\sqrt{\sigma_s}}{\T}$ at $T=\T$ to the data given in 
Table 1 of \cite{bali} we now fix the value  of $\g$ and obtain\footnote{Since $\g$ depends 
on a number of colors, we have to adjust 
its value to $SU(2)$.} 

\begin{equation}\label{fit}
\frac{\sqrt{\sigma_s}}{\T}=1.44\,\frac{T}{\T}
\exp\Bigl\lbrace\frac{1}{2}\Bigl(\frac{\T}{T}\Bigr)^2-\frac{1}{2}\Bigr\rbrace
\,.
\end{equation}
In Fig.4 we have plotted $\tfrac{\sqrt{\sigma_s}}{\T}$ against $\tfrac{T}{\T}$. 
%
\begin{figure}[ht]
\begin{center}
\includegraphics[width=6.75 cm]{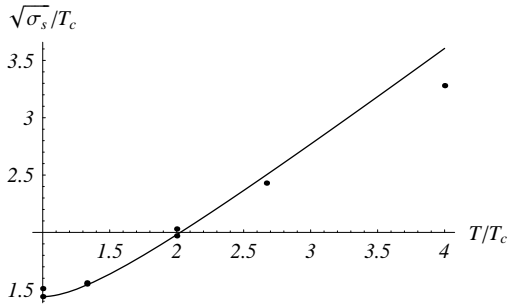}
\caption{\small{Square root of the spatial string tension versus temperature in units of $\T$. The dots denote the data from \cite{bali}.
}}
\end{center}
\end{figure}
We find that the temperature dependence of the 
spatial string tension is in good agreement with the lattice data in the region $1\leq\tfrac{T}{\T}\lesssim 2.5$. From this 
point of view it can be thought of as a description 
of the spatial string tension in the high temperature phase near the phase 
transition point. Note that the temperature dependence of 
the spatial string tension at high temperatures is determined by 
the $\beta$-function of gauge theory. In particular the logarithmic dependence is seen
in the data at high temperatures.\footnote{For a discussion of this issue on the lattice, see, e.g., \cite{bali,karsch} and references therein.}
The model \eqref{metric} cannot reproduce logarithmic dependences
associated with the small coupling running. Instead, it provides a complementary description in the 
strong coupling regime that is in accord with the ideas on the AdS/CFT correspondence.

\section{Concluding Comment}
In conclusion, we would like to emphasize again  that we have {\it not} introduced any {\it new} parameter to describe the critical 
temperature as well as the temperature dependence of the spatial string tension. The only dimensionful parameter of our model is $c$. 
It appears as the Regge parameter at zero temperature. Moreover, the physical string tension at zero temperature is proportional to $c$ 
with the coefficient of proportionality fixed from the linear term of the Cornell potential \cite{az}. The temperature of the phase transition
appears then calculable. Note a crucial factor of $1/2\pi^2$ which explains, in the framework considered, the lower scale
of the critical temperature, $\T^2\sim0.04\,\text{ GeV\,}^2$, than $c\sim0.9\,\text{GeV\,}^2$. Moreover, the temperature dependence 
of the spatial string tension is very soft below $\T$ and sharp above $\T$. This feature of the data  is also explained by the model. To 
summarize, the AdS/QCD correspondence provides a quantitative framework for temperatures of order $\T$ which works with 
about $10 \%$ accuracy. 
\vspace{.25cm}

{\bf Acknowledgments}

\vspace{.25cm}
O.A. would like to thank G. Lopes Cardoso and P. Weisz for useful discussions. The  work  of O.A. was supported in part by 
Max-Planck-Gesellschaft and Russian Basic Research Foundation Grant 05-02-16486. O.A. also 
thanks D. L\" ust for hospitality at the Heisenberg Institut, where a main portion of this work was completed.



\small
\begin{thebibliography}{99}
\bibitem{pseudo}
C. Borg, Nucl.Phys. B261 (1985) 455;\\
E. Manousakis and J. Polonyi, Phys.Rev.Lett. 58 (1987) 847.
\bibitem{bali}
G. Bali, J. Fingberg, U.M. Heller, F. Karsch, and K. Schilling, Phys.Rev.Lett. 71 (1993) 3059.
\bibitem{karsch}
F. Karsch, E. Laermann, and M. L\" utgemeier, Phys.Lett.B 346 (1995) 94.
\bibitem{az}
O. Andreev and V.I. Zakharov, Heavy-quark potentials and AdS/QCD, hep-ph/0604204.
\bibitem{witt}
E. Witten, Adv.Theor.Math.Phys. 2 (1998) 505.
\bibitem{mets}
R.R. Metsaev, IIB supergravity and various aspects of light cone formalism in AdS space-time, hep-th/0002008;\\
A. Karch, E. Katz, D.T. Son, and M.A. Stephanov, Linear Confinement and AdS/QCD, hep-ph/0602229.
\bibitem{oa}
O. Andreev,  Phys.Rev.D 73 (2006) 107901.
\bibitem{Tloops}
The following is an incomplete list: \\
S.-J. Rey, S. Theisen, and J.-T. Yee,  Nucl.Phys.B 527 (1998) 171;\\
A. Brandhuber, N. Itzhaki, J. Sonnenschein, and S. Yankielowicz, JHEP 9806 (1998) 001;\\
D.J. Gross and H. Ooguri, Phys.Rev.D 58 (1998) 106002;\\
H. Dorn and H.J. Otto, JHEP 9809 (1998) 021;\\
S. Naik, Phys.Lett.B 464 (1999) 73;\\
Y. Kinar, E. Schreiber, and J. Sonnenschein, Nucl.Phys.B 566 (2000) 103;\\
F. Bigazzi, A.L. Cotrone, L. Martucci, and L.A. Pando Zayas, Phys.Rev.D 71 (2005) 066002;\\
 S.A. Hartnoll and S. P. Kumar, Multiply wound Polyakov loops at strong coupling, hep-th/0603190.
\bibitem{gr}
I.S. Gradshteyn and I.M. Ryzhik, Table of Integrals, Series, and Products, Academic Press, 1994. 
\end{thebibliography}
\end{document}